\documentclass[reprint,
bibnotes,
amsmath,amssymb,
aps,
prl,
]{revtex4-2}

\usepackage{dcolumn}
\usepackage{bm}
\usepackage{url}
\usepackage[dvipdfmx]{hyperref}
\usepackage[dvipdfmx]{graphicx}

\begin{document}

\title{
Effects of Electron Correlations and Chemical Pressures\\
on Superconductivity of $\beta^{\prime\prime}$-type Organic Compounds
}

\author{
Shusaku Imajo$^{1,2}$}
\thanks{imajo@issp.u-tokyo.ac.jp}
\author{
Hiroki Akutsu$^1$,
Akane Akutsu-Sato,
Alexander L. Morritt$^3$, 
Lee Martin$^3$, and
Yasuhiro Nakazawa$^1$
}

\affiliation{
$^1$Graduate School of Science, Osaka University, Toyonaka, Osaka 560-0043, Japan\\
$^2$The Institute for Solid State Physics, University of Tokyo, Kashiwa, Chiba 277-8581, Japan\\
$^3$School of Science and Technology, Nottingham Trent University, Clifton Lane, Nottingham NG11 8NS, United Kingdom
}
\date{\today}

\begin{abstract}
 We investigate low-temperature electronic states of the series of organic conductors $\beta^{\prime\prime}$-(BEDT-TTF)$_4$[(H$_3$O)M(C$_2$O$_4$)$_3$]G, where BEDT-TTF is bis(ethylenedithio)tetrathiafulvalene, and M and G represent trivalent metal ions and guest organic molecules, respectively.
Our structural analyses reveal that the replacement of M and G give rise to systematic change in the cell parameters, especially in the $b$-axis length, which has positive correlation with the superconducting transition temperature $T_{\rm c}$.
Analyses of temperature and magnetic field dependences of the electrical resistance including the Shubnikov-de Haas oscillations elucidates that the variation of charge disproportionation, effective mass and the number of itinerant carriers, can be systematically explained by the change of the $b$-axis length.
The changes of the transfer integrals induced by stretching/compressing the $b$-axis are confirmed by the band calculation.
We discuss that electron correlations in quarter-filled electronic bands lead to charge disproportionation and the possibility of a novel pairing mechanism of superconductivity mediated by charge degrees of freedom. 
\end{abstract}

\maketitle

 Superconductivity dominated by electron correlations often appears in nearly half-filled electronic bands, where an antiferromagnetic Mott insulating state is easily formed by on-site Coulomb repulsion $U$.
In such cases, the superconductive regions are located in close proximity to the magnetic Mott phases in electronic phase diagrams\cite{1,2,3}.
Therefore, unconventional pairing related to magnetic spin fluctuations has been suggested to provide an understanding of the mechanisms of the superconductivity.
Indeed, highest critical temperatures $T_{\rm c}$ are normally observed on the verge of the magnetic phases because the quantum fluctuation coming from the spin degree of freedom is strongly enhanced.
High-$T_{\rm c}$ cuprates\cite{1}, heavy fermion superconductors\cite{2} and dimer-Mott type organic superconductors\cite{3} have been discussed as such candidates of spin-fluctuation-mediated superconductors.
Other degrees of freedom, such as orbital (multipole)\cite{4,5} and charge\cite{6}, have been also proposed as origins of the pairing in some superconductors, $e.g.$ iron-based compounds, cage compound, $etc$.
However, it is difficult to examine the relationship between the quantum degrees of freedom and superconductivity because the number of such exotic superconductors is rather small and the pairing states are often in complicated situations due to the coexistence/competition of some degrees of freedom.

\begin{figure}
\begin{center}
\includegraphics[width=\hsize,clip]{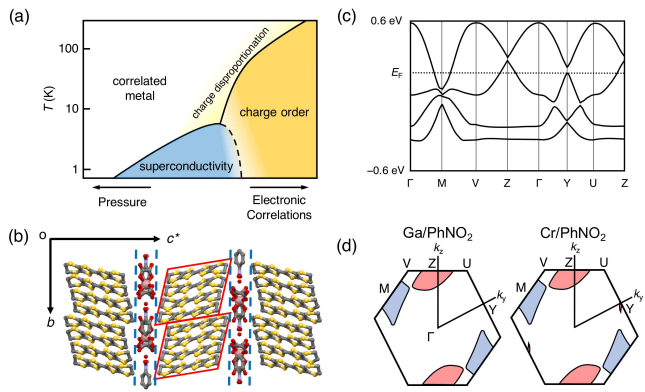}
\end{center}
\caption{(Color online)
(a) A schematic phase diagram of 1/4-filling $\beta^{\prime\prime}$-type organic conductors that the superconducting phase is located at the proximity to the charge ordered phase.
(b) Interlayer packing structure of the present salts $\beta^{\prime\prime}$-(BEDT-TTF)$_4$[(H$_3$O)M(C$_2$O$_4$)$_3$]G.
The red rhombi indicate tetramers of the BEDT-TTF molecules.
The blue dashed lines denote the region of the insulating layer composed of [(H$_3$O)M(C$_2$O$_4$)$_3$]$^{2-}$ and G.
(c) Band structure of one of the salts, $\beta^{\prime\prime}$-(BEDT-TTF)$_4$[(H$_3$O)Cr(C$_2$O$_4$)$_3$]PhNO$_2$.
(d) The Fermi surface and first Brillouin zone of $\beta^{\prime\prime}$-(BEDT-TTF)$_4$[(H$_3$O)Ga(C$_2$O$_4$)$_3$]PhNO$_2$ and $\beta^{\prime\prime}$-(BEDT-TTF)$_4$[(H$_3$O)Cr(C$_2$O$_4$)$_3$]PhNO$_2$ derived from the band calculation.
}
\label{fig1}
\end{figure}

 Recently, $\beta^{\prime\prime}$-type organic charge-transfer salts consisting of BEDT-TTF molecules with counter anions and guest solvent molecules have drawn extensive attention because they are expected to have a novel Cooper pairing mechanism related to charge degrees of freedom\cite{6,7,8,9,10,11,12,12p5,13,14}.
The critical temperatures $T_{\rm c}$ of some $\beta^{\prime\prime}$-salts are relatively high around 7-8~K, which is almost comparable to the $T_{\rm c}$ of well-known higher-$T_{\rm c}$ dimer-Mott organic superconductors, $\kappa$-(BEDT-TTF)$_2$X.
Also, the strong-coupling superconductivity with the large superconducting energy gap $\Delta$$\sim$2.5$k_{\rm B}$$T_{\rm c}$ is expected for $\beta^{\prime\prime}$-(BEDT-TTF)$_4$[(H$_3$O)Ga(C$_2$O$_4$)$_3$]PhNO$_2$\cite{17,18}.
As is shown in Fig.~\ref{fig1}(a), the universal phase diagram of the $\beta^{\prime\prime}$-type salts proposed in the earlier works\cite{11,12,12p5,13,14} claims that the superconducting phase is adjacent to the charge ordered (CO) phase caused by inter-site Coulomb repulsion $V$.
It implies that the CO phase appears as an underlying phase for the emerging superconductive phase.
Figure~\ref{fig1}(b) displays a crystal structure of one such example, a series of the $\beta^{\prime\prime}$-type salts, $\beta^{\prime\prime}$-(BEDT-TTF)$_4$[(H$_3$O)M(C$_2$O$_4$)$_3$]G (abbreviated as M/G in this report), composed of BEDT-TTF (bis(ethylenedithio)tetrathiafulvalene), tris(oxalato)metallate ions M(C$_2$O$_4$)${_3}^{3-}$ and guest molecules G.
The $\beta^{\prime\prime}$-type salts are classified as non-dimeric type organic compounds due to the absence of dimerization of the BEDT-TTF molecules.
As the non-dimeric feature and the valence (BEDT-TTF)$^{0.5+}$ make the energy band quarter filled, magnetic degrees of freedom are expected to be quenched in uniform CO states.
Some earlier works\cite{16p5,15,16} suggest that $T_{\rm c}$ of the series also exhibits chemical pressure effect associated with the universal phase diagram, depending on volume\cite{15,16} and length of the guest molecules\cite{16p5} of the salts.
In this work, to elucidate the electronic states of the $\beta^{\prime\prime}$-type salts in more detail, we investigate the low-temperature electronic properties by means of electrical transport measurements with systematic changes of the electronic states through the chemical substitutions.
From the systematic analyses of Shubnikov–de Haas (SdH) oscillations, crystal axes and band calculations, we clarify the effect of chemical substitution on the electronic state and the relationship between the superconductivity and physical parameters.

 Single crystals of the compounds displayed in Table~\ref{tab1} were synthesized by electrochemical oxidation methods.
The crystal structures and axes of the salts are determined by X-ray diffraction at 298~K.
The out-of-plane resistance was measured by a standard four-terminal AC method with current along $c^*$-axis.
Since the crystals harvested from the same batch do not have large sample dependences, each data point in the figures of this paper is obtained by the crystals from one batch for one composition.
Four gold wires were attached by carbon paint on both plane surfaces of single crystals.
The transport measurements were performed in a $^3$He-refrigerator with a 15~T superconducting magnet and a 14~T Physical Property Measurement System (Quantum Design).
\begin{table}
\begin{center}
\begin{tabular}{|c|c|c|m{5mm}|c|c|c|}\hline
\multicolumn{7}{|c|}{$\beta^{\prime\prime}$-(BEDT-TTF)$_4$[(H$_3$O)M(C$_2$O$_4$)$_3$]G}\\ \hline
M & G & $T_{\rm c}$ & & M & G & $T_{\rm c}$\\ \hline \hline
Fe & C$_6$H$_5$CN & 7.2~K & & Ga & 2-C$_5$H$_4$ClN & 1.7~K\\
Ga & C$_6$H$_5$NO$_2$ & 7.0~K & & Ga & C$_6$H$_5$Cl & no SC\\
Ga & DMF & 4.5~K & & Rh & C$_6$H$_5$Cl & no SC\\
Ga & C$_6$H$_5$Br & 3.0~K & & Ru & C$_6$H$_5$Cl & no SC\\
Rh & C$_6$H$_5$Br & 2.9~K & & Ga & C$_6$H$_5$F & no SC\\
Ru & C$_6$H$_5$Br & 2.8~K & & Rh & C$_6$H$_5$F & no SC\\
Ga & C$_5$H$_5$N & 2.5~K & & Ru & C$_6$H$_5$F & no SC\\
Ga & C$_6$H$_5$I & 2.4~K & & Ga & CH$_2$HCl$_2$ & no SC\\ \hline
\end{tabular}
\caption{The metal ion M and guest molecule G of the salts measured in the present study and their critical temperature $T_{\rm c}$.}
\label{tab1}
\end{center}
\end{table}

 As is shown in Fig.~\ref{fig1}(b), the series of $\beta^{\prime\prime}$-type salts have a layered structure of alternately stacked BEDT-TTF donor layers and counter anions and guest solvent molecules layers.
The non-dimerized arrangement of the donors make the electronic state quarter-filled, since the four BEDT-TTF molecules form charge transfer salts with the divalent counter anions [(H$_3$O)M(C$_2$O$_4$)$_3$]$^2-$.
Additionally, the counter anions layer has a neutral solvent molecule, G, aligned along the $b$-axis with commensurate periodicities.
The compounds have weak tetramerization (red rhombus) as shown in Fig.~\ref{fig1}(b), and therefore, the band structure and Fermi surface is semi-metallic as is depicted in Figs.~\ref{fig1}(c),(d).

\begin{figure}
\begin{center}
\includegraphics[width=\hsize,clip]{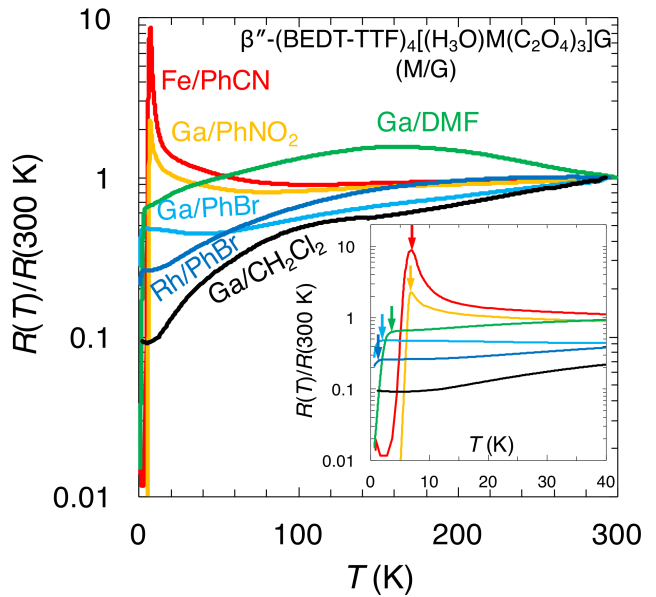}
\end{center}
\caption{(Color online)
Temperature dependences of the scaled electric resistance measured on some salts.
Inset shows the expanded graph below 40~K.
The arrows indicate the superconducting transition temperatures $T_{\rm c}$.
}
\label{fig2}
\end{figure}

 Figure~\ref{fig2} presents the temperature dependences of the interlayer electrical resistance for several $\beta^{\prime\prime}$-compounds in a semi-logarithmic plot.
The temperature dependences of the salts including the other salts also measured in this work [see Table~\ref{tab1}] qualitatively reproduced the results of the previous works\cite{15,16,17,19,20,21,22,23,24}.
The temperature dependences of the electrical resistance above 100 K is moderate in all salts in the plot.
The higher-$T_{\rm c}$ salts, such as Fe/PhCN and Ga/PhNO$_2$, show weak semiconducting behavior down to 10~K and they show abrupt increase of the resistance below about 10~K, while the lower-$T_{\rm c}$ salts and no-SC salts such as Rh/PhBr and Ga/CH$_2$Cl$_2$ show only weak semiconducting behavior at lower temperatures\cite{15,16,17,19,20,21,22,23,24}.
This feature is shown in the inset of Fig.~\ref{fig2}.
Since the gradual increase of the resistance upon cooling is found to have a logarithmic temperature dependence $\sim$log$T$, the origin is discussed from a viewpoint of weak localization which is the quantum interference in disordered electronic systems in the previous report\cite{24}.
On the other hand, the insulating behavior below 10~K is observed only in the higher-$T_{\rm c}$ salts.
It is considered that the mobility of itinerant carriers is suppressed by developments of charge disproportionation (CD)\cite{7,8,10}.
In the case of Ga/PhNO$_2$, the NMR and EPR studies\cite{7,8,10} reveal that the CD appears at very close to $T_{\rm c}$ with large charge imbalance $\Delta \rho$ $\sim$0.5.
The transition temperature to the CD state $T_{\rm CD}$ is about 8.5~K, which is just above $T_{\rm c}$~$\sim$7.0 K.
It means that the superconductivity occurs in the CD state whose nature is electrically semiconducting.
Also, the CD state has itinerant carriers to form the superconductivity, implying the existence of the Fermi surface even in the CD state.
The higher-$T_{\rm c}$ salts have enough strong electron correlation to make the large CD, which is the precursor of the rigid charge ordered state, in contrast to the case of the lower-$T_{\rm c}$ salts and no-SC salts.
The enhancement of $T_{\rm c}$ with increasing the electron correlation suggests that the superconductivity may be promoted by the charge instability evolved around the verge of the charge ordered phase.

\begin{figure}
\begin{center}
\includegraphics[width=\hsize,clip]{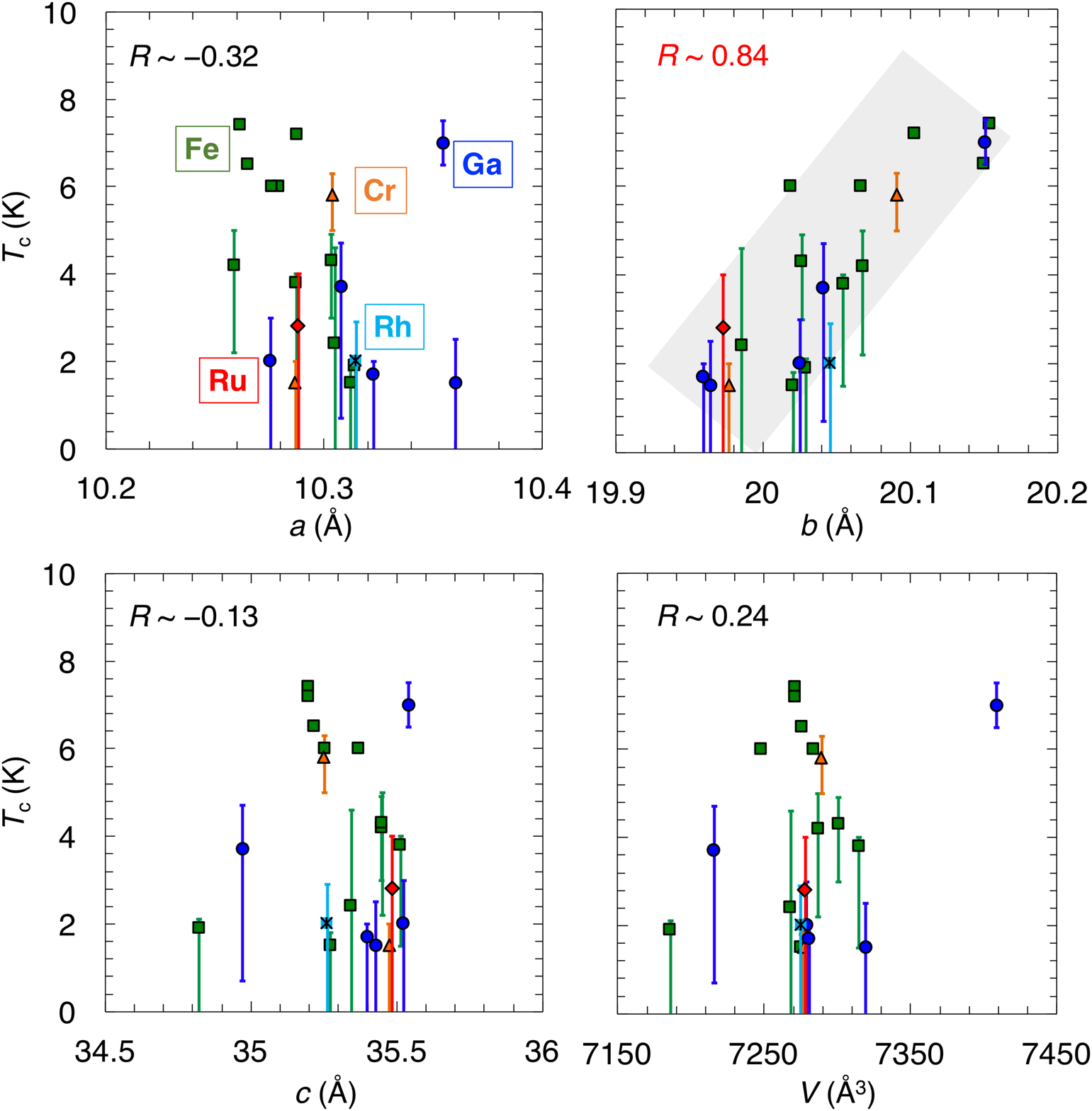}
\end{center}
\caption{(Color online)
Superconducting critical temperatures $T_{\rm c}$ vs. lattice constants of the $\beta^{\prime\prime}$-type salts and correlation coefficient of the plots $R$.
It includes the data reported in the previous works\cite{15,16,19,20,21,22,23,24}.
The red, blue, green, orange, cyan symbols represent the center metal ion M of tris(oxalato)metallate as Ru, Ga, Fe, Cr, Rh, respectively.
$T_{\rm c}$ of the salts measured in this work is defined by the intersection of the temperature dependence of the resistance of the normal and superconducting states.
The bars indicate the onset and zero-resistivity temperatures.
The symbols without bars are data taken from previous works for which we cannot determine bars.
The error bars of the lattice constants are smaller than the size of the symbols.
}
\label{fig3}
\end{figure}

\begin{figure}
\begin{center}
\includegraphics[width=\hsize,clip]{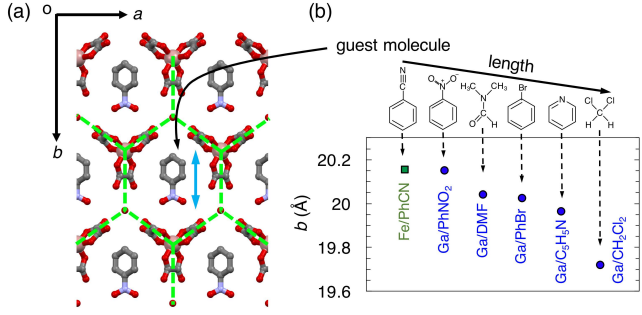}
\end{center}
\caption{(Color online)
(a) The arrangement of the guest molecules G in the hexagonal vacancies enclosed by the tris(oxalato)metallate and oxonium ions.
(b) Schematic comparison of the length of the guest molecules G and the $b$-axis length of the $\beta^{\prime\prime}$-type salts containing the respective guest molecules.
}
\label{fig4}
\end{figure}

 In order to discuss the difference in $T_{\rm c}$, we now turn to the chemical pressure effect against the electronic system through cation and guest molecule substitution.
Since the replacement of the trivalent metal ion M and the solvent molecules G does not modify the band filling, but induces local volume change, the substitution must work as the change of the energy dispersion and bandwidth of the $\pi$-electrons.
It should also be mentioned that introduction of magnetic ions like Fe$^{3+}$ (S = 5/2) and Cr$^{3+}$ (S = 3/2) does not exert a strong influence on the $\pi$-electrons.
Although the interaction between the insulating and conducting layers known as $\pi$-d interaction\cite{26} plays an important role for the electronic state in some organics, the $\pi$-d interaction in the present class is negligible because the metal ion M is located in the depth of the tris(oxalato)metallate molecules.
It is indeed confirmed by the absence of a difference of $T_{\rm c}$ between Fe/G (S = 5/2) salts and Ga/G (non-magnetic) salts with the same G\cite{15,16}.
To evaluate the chemical pressure effect, $T_{\rm c}$ is plotted as functions of the respective lattice constants ($a$-, $b$-, $c$-axes and volume) at 298 K in Fig.~\ref{fig3}.
The data reported in previous literature\cite{15,16,19,20,21,22,23,24,27,28,29} are also included.
The correlation coefficient $R$ for each plot unveils that only the length of the $b$-axis has the strong correlation with $T_{\rm c}$, $\mid$$R$$\mid$~$\sim$0.84, whereas no obvious correlations $\mid$$R$$\mid$~$<$~0.3 are observed in the plots against the $c$-axes and volume.
The weak negative value $R$~${\sim}$−0.32 for the $a$-axis should originate from the Poisson effect in the $a$-$b$ conducting planes, and therefore, it is reasonable to consider that the $b$-axis length gives the most dominant effect on $T_{\rm c}$.
In the counter anions layers, the $b$-axis is the direction toward which the guest molecules orient as represented in Fig.~\ref{fig4}(a).
It indicates that the length of the guest molecules is exactly the main cause of the substitution effect.
As we can find in Table~\ref{tab1}, the salts with the longer guest molecules in Fig.~\ref{fig4}(b) like PhCN, indeed, show higher-$T_{\rm c}$ whereas salts with the shorter guest molecules like CH$_2$Cl$_2$ do not superconduct.
Namely, the electronic states of the present class can be controlled by the $b$-axis length through the size of the guest molecules as expected in the previous work\cite{16p5}.
Hereafter, we use the $b$-axis length to evaluate the electronic states.

\begin{figure}
\begin{center}
\includegraphics[width=\hsize,clip]{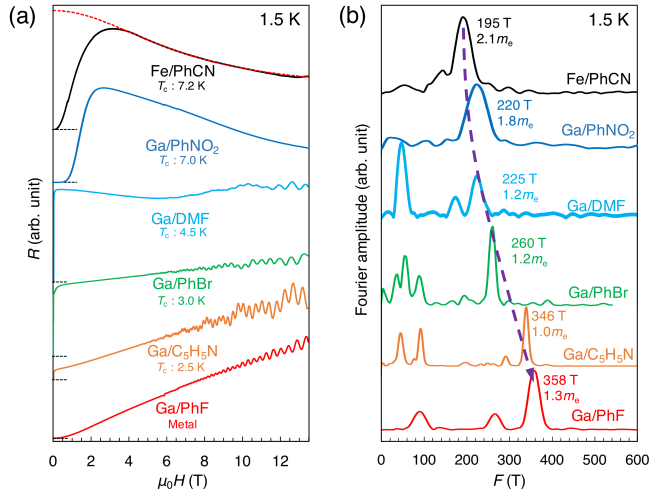}
\end{center}
\caption{(Color online)
(a) Magnetoresistance of some $\beta^{\prime\prime}$-salts at 1.5~K in magnetic fields perpendicularly applied to the conducting plane.
The data have been offset for clarity.
The dashed red line denotes the theoretical calculation expected by the weak localization model\cite{31}.
(b) The Fourier spectra of the Shubnikov–de Haas oscillations of the samples shown in Fig.~\ref{fig5}(a).
The values next to the peaks indicate the frequency (upper) and effective mass (lower) of the respective SdH oscillations.
}
\label{fig5}
\end{figure}

\begin{figure}
\begin{center}
\includegraphics[width=\hsize,clip]{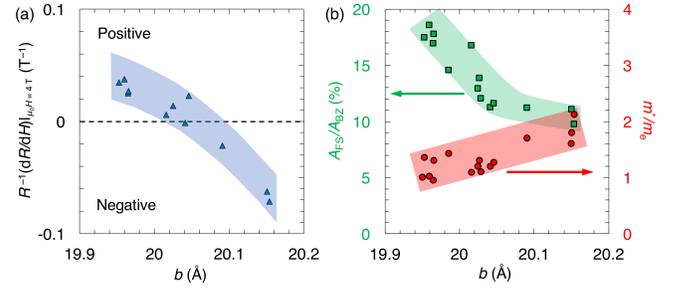}
\end{center}
\caption{(Color online)
(a) The $b$-axis length dependence of the normalized slope of the magnetoresistance (d$R$/d$T$)/$R$ at 4~T.
The sign of the value corresponds to the sign of the magnetoresistance.
(b) Percentage of the cross-sectional area of the electron pocket and the first Brillouin zone (green, the left axis) and the effective mass (red, the right axis) as a function of the $b$-axis length.
}
\label{fig6}
\end{figure}

 For the discussion of the itinerant carriers, magnetoresistance of some samples at 1.5~K in perpendicular magnetic fields is shown in Fig.~\ref{fig5}(a) in descending order of the $b$-axis length from upper to lower.
The abrupt increase of the resistance below 1 T comes from the suppression of the superconductivity.
Whilst we see the presence of negative magnetoresistance of the higher-$T_{\rm c}$ salts, only positive magnetoresistance is observed in the lower-$T_{\rm c}$ and metallic salts.
Figure~\ref{fig6}(a) shows the normalized slope of the magnetoresistance (d$R$/d$T$)/$R$ at 4~T so as to evaluate the change of the magnetoresistance.
The magnitude of the negative magnetoresistance is suppressed in compounds with shorter $b$-axis lengths and it finally converts into a positive magnetoresistance.
The negative magnetoresistance is discussed in other $\beta^{\prime\prime}$-type salts in terms of Anderson localization with charge density wave (CDW)\cite{11,30}.
They suggest that the presence of the CDW can introduce scattering centers, which produce weak localization.
Similar to the case of these references, the present result is also explained by the negative magnetoresistance based upon the weak localization proposed by Fukuyama and Yoshida\cite{31} as displayed by the red dashed line of Fe/PhCN in Fig.~\ref{fig5}(a) even though the ratio of the charge disproportionation may be rather large as compared with CDWs.
The result also agrees with the logarithmic temperature dependence of electrical resistance discussed from the viewpoint of the weak localization as mentioned above\cite{24}.
The $b$-axis length dependence of the magnetoresistance suggests that the higher-$T_{\rm c}$ salts have larger electron correlation to make the CD than the lower-$T_{\rm c}$ salts have.
In other words, the electron correlation of the present class is also related to the $b$-axis length, similar to $T_{\rm c}$.

 In the magnetoresistance curves at lower temperatures, Shubnikov–de Haas (SdH) oscillations are observed in all salts except for the M=Ru cases.
The higher-$T_{\rm c}$ salts do not have large oscillations as compared with those of lower-$T_{\rm c}$ ones, but, tiny modulation of the resistivity certainly appears in higher-field region above 10~T.
The smaller amplitude of the oscillation of the higher-$T_{\rm c}$ salts may be caused by the enhanced scattering rate owing to the coexistence with the larger CD.
Nonetheless, the observation of the SdH oscillations indicates the existence of the Fermi surface even in the CD state for the higher-$T_{\rm c}$ salts.
Although it is still an open question how the CD state coexists with the conducting part, the electrons in the Fermi surface allows to the semiconducting CD state to show the superconductivity.
The oscillations of the magnetoresistance can be transformed into the Fourier transform spectrum, which shows peaks at the frequency of the SdH oscillations as seen in Fig.~\ref{fig5}(b).
The higher-$T_{\rm c}$ salts show only one peak that corresponds to the electron and hole pockets of which cross-sectional area of Fermi surface is the same as expected in the simple compensated metal.
This is consistent with the results of the previous SdH study by Bangura $et~al$.\cite{16,17} and the calculated Fermi surface reported in Ref.~\cite{17}.
In contrast to this, the lower-$T_{\rm c}$ salts have some extra peaks\cite{15,27,28}.
In the Brillouin zone of the middle-$T_{\rm c}$ salt Cr/PhNO$_2$ ($T_{\rm c}$:~${\sim}$5 K)\cite{16} as introduced in Fig.~\ref{fig1}(e), we can find that there is an additional tiny hole pocket around Y point whereas the higher-$T_{\rm c}$ salt Ga/PhNO$_2$ does not show.
It means that the topology of the Fermi surface changes by the Lifshitz transition when decreasing the $b$-axis length.
When the bandwidth is stretched by the increase of transfer integral concomitant with the contraction of cell parameters, the additional hole pocket can appear at the Y point and all of the Fermi pockets becomes larger.
Hence, the lower-$T_{\rm c}$ salts have the larger Fermi pockets with the additional small Fermi pocket.
The emergence of the additional Fermi pocket gives some peaks in the spectra due to the split of the hole pockets and the formation of new magnetic breakdown orbits.
For simplicity of the discussion, we hereafter focus on the electron pocket at the M point because it does not split in the scope of the present chemical substitution.
The dotted line in Fig.~\ref{fig5}(b) is a guide for the eye to emphasize the peak shift of the electron pocket.
In Fig.~\ref{fig6}(b), we present the ratio of cross-sectional area of the electron pocket and Brillouin zone $A_{\rm FS}$/$A_{\rm BZ}$ and effective mass of the electron as $m^*$/$m_{\rm e}$ against the $b$-axis length including the data reported in previous work\cite{15,16,27,28}.
The effective mass $m^*$ is derived from the temperature dependence of the amplitude of each peak by using the Lifshitz-Kosevich (LK) formula.
Similar to the case of the $T_{\rm c}$ and the magnetoresistance, these parameters also exhibit the clear correlation with the $b$-axis length.
When decreasing the number of conducting carriers, the area of the Fermi surface in the Brillouin zone becomes smaller.
Therefore, the diminution of the ratio $A_{\rm FS}$/$A_{\rm BZ}$ discussed here demonstrates the gradual decrease of the number of conducting carries while the effective electron mass is enhanced in the higher-$T_{\rm c}$ salts as shown in Fig.~\ref{fig6}(b).
The extension of the $b$-axis length due to the replacement by larger guest molecules reduces the transfer integrals in the molecular stacking.
Since the band structure of the present salts is semi-metallic, this structural change leads the shrinkage of the size of the Fermi pockets in semi-metallic bands accompanied by the reduction of the bandwidth $W$.
The declining $W$ in the salts with larger $b$-axis should give the augmentation of the electron correlations because of the increase of the $V$/$W$ ratios.
Namely, the electron mass enhancement appears and the growth of charge disproportionation also occurs in the larger $b$-axis compounds.
Hence, these results agree with the electronic phase diagram presented in Fig.~\ref{fig1}(a).

\begin{figure}
\begin{center}
\includegraphics[width=\hsize,clip]{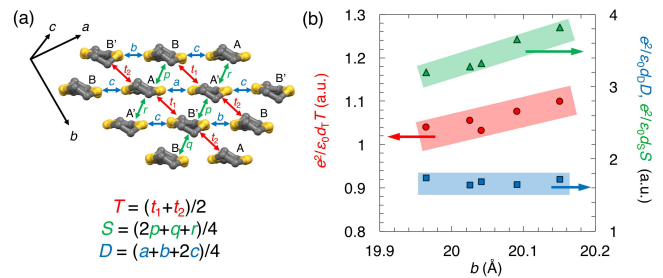}
\end{center}
\caption{(Color online)
(a) The placement of the transfer integrals between BEDT-TTF molecules in the conducting layer.
The transfer integrals are classified by their directions into three types, $T$, $S$, and $D$.
The lower formulae represent the definition of the averaged three transfer integrals.
(b) The $b$-axis dependences of the electron correlation along each direction.
The evaluation of the electron correlation $V$/$W$ is described in the main text.
}
\label{fig7}
\end{figure}

 In order to confirm these experimental speculations, we perform the band calculation by extended-H$\rm{\ddot{u}}$ckel tight binding method\cite{32} with a unit cell transformation from the $C$-centered lattice ($x$, $y$, $z$) to the primitive lattice ($z$, 0.5$x$+0.5$y$, −$x$)\cite{16}.
In the present $\beta^{\prime\prime}$-family, there are eight types of transfer integrals ($t_{\rm 1}$, $t_{\rm 2}$, $p$, $q$, $r$, $a$, $b$, $c$) between the BEDT-TTF molecules as seen in Fig.~\ref{fig7}(a).
We here introduce the averaged transfer integrals, transverse $T$ (=($t_{\rm 1}$+$t_{\rm 2}$)/2), stacking $S$ (=(2$p$+$q$+$r$)/4) and diagonal $D$ (=($a$+$b$+2$c$)/4) according to the directions and the number of bonds.
In the case of the 1/2-filling Mott systems, the electron correlation is described as $U/W$, where the on-site Coulomb repulsion $U$ is independent of inter-site distance.
However, the inter-site Coulomb repulsion $V$ of the electron correlation in the present 1/4-filling system $V$/$W$ depends on the inter-site distance.
Since $V$ is inversely proportional to the distance between molecules and $W$ is in proportion to the transfer integrals, the electron correlation can be evaluated as the dimensionless quantity $V/W\sim e^{\rm 2}/\varepsilon_{\rm 0}Td_{\rm T}$, $e^{\rm 2}/\varepsilon_{\rm 0}Sd_{\rm S}$ and $e^{\rm 2}/\varepsilon_{\rm 0}Dd_{\rm D}$ by the electric charge $e$, the electric constant $\varepsilon_{\rm 0}$, the transfer integrals and the average distance between the molecules along the respective directions $d_{\rm T}$, $d_{\rm S}$ and $d_{\rm D}$.
Figure~\ref{fig7}(b) displays the $b$-axis length dependence of the electron correlation along each direction.
The diagonal electron correlation $e^{\rm 2}/\varepsilon_{\rm 0}Dd_{\rm D}$ seems to be independent on the $b$-axis length whereas the other two $e^{\rm 2}/\varepsilon_{\rm 0}Td_{\rm T}$ and $e^{\rm 2}/\varepsilon_{\rm 0}Sd_{\rm S}$ show positive correlation with the $b$-axis length.
It is natural because the $b$-axis is the direction nearly perpendicular to the $D$ direction while the others possess the parallel component to the $b$-axis.
The positive correlation of the two means that the elongation increases the electron correlation because of the reduction of the overlap of the wave function along the $b$-axis.
It confirms that the band calculation perfectly demonstrates the above discussion.
Hence, our results totally indicate that the transfer integrals along the $b$-axis play an important role for the electronic state.
The CD is considered as the consequence of the charge instability near the charge ordered phase by increasing the electron correlation $V$/$W$.

\begin{figure}
\begin{center}
\includegraphics[width=\hsize,clip]{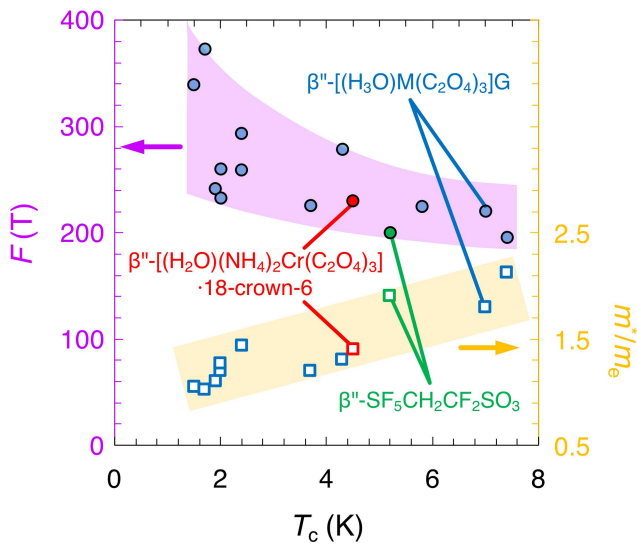}
\end{center}
\caption{(Color online)
The comprehensive plot of the frequency of the Shubnikov–de Haas oscillations $F$ (left) and the effective mass $m^*$/$m_{\rm e}$ (right) vs. $T_{\rm c}$ for 1/4-filled $\beta^{\prime\prime}$-type organic salts.
The colors of the symbols represent the type of the counter anions of the $\beta^{\prime\prime}$-type salts as [(H$_3$O)M(C$_2$O$_4$)$_3$]G (blue), SF$_5$CH$_2$CF$_2$SO$_3$(green) and [(H$_2$O)(NH$_4$)$_2$Cr(C$_2$O$_4$)$_3$]$\cdot$18-crown-6 (red), respectively.
While the circles denote the data of the frequency, the boxes indicate those of the effective mass.
}
\label{fig8}
\end{figure}

 Finally, we mention the systematic relation between $T_{\rm c}$, the effective mass, and the cross-section of the Fermi surface in the case of other 1/4-filled $\beta^{\prime\prime}$-type organic superconductors.
In Fig.~\ref{fig8}, we present the plot of $m^*$ and the SdH frequency $F$ of $\beta^{\prime\prime}$-(BEDT-TTF)$_2$SF$_5$CH$_2$CF$_2$SO$_3$ ($T_{\rm c}$~:~$\sim$5.2~K, $m^*$~:~$\sim$1.9$m_{\rm e}$, $F$~:~$\sim$200~T)\cite{33} and $\beta^{\prime\prime}$-(BEDT-TTF)$_2$[(H$_2$O)(NH$_4$)$_2$Cr(C$_2$O$_4$)$_3$]$\cdot$18-crown-6 ($T_{\rm c}$~:~$\sim$4.5~K, $m^*$~:~$\sim$1.4$m_{\rm e}$, $F$~:~$\sim$230~T)\cite{34,35} with the results of the present salts.
It is clearly found that the parameters of the two salts fall in the correlation derived from our results.
Quite recently, optical measurements\cite{13,14} point out that the superconductivity found in $\beta^{\prime\prime}$-(BEDT-TTF)$_2$SF$_5$CH$_2$CF$_2$SO$_3$ also coexists with the charge order similar to the present case.
These facts suggest that the $\beta^{\prime\prime}$-type organics can be universally discussed in a generic phase diagram with the electron correlation like Fig.~\ref{fig1}(a) and the superconductivity observed in these $\beta^{\prime\prime}$-type salts should share the same pairing mechanism of the superconductivity.
In future work, the symmetry of the superconducting energy gap should be examined for the comprehension of the detailed mechanism of how the superconductivity is formed by the charge fluctuation.

 We reported the electrical transport properties of the organic conductors $\beta^{\prime\prime}$-(BEDT-TTF)$_4$[(H$_3$O)M(C$_2$O$_4$)$_3$]G with chemical substitution.
From the results of the temperature and magnetic field dependences, we find that the semiconducting behavior and negative magnetoresistance observed in the higher-$T_{\rm c}$ salts originate from the charge disproportionation caused by the stronger electron correlation.
We also find that the superconducting transition temperatures are influenced mainly by the $b$-axis length depending on the length of the guest molecules because the $b$-axis corresponds to the direction of the orientation of the guest molecules in the counter layer.
From the analyses of the Shubnikov–de Haas oscillations, it was found that the effective mass and the number of carriers are influenced by the $b$-axis length.
The band calculation also demonstrates the experimental results that the electron correlation reflects the transfer integrals along the $b$-axis direction.
The increase of $T_{\rm c}$ accompanied by the enhancement of the CD with lengthening the $b$-axis length suggests that the superconductivity should be mediated by the charge fluctuation evolved at the phase boundary of the charge ordered phase.
Moreover, we find that the present discussion of the electronic state can be applied to the other $\beta^{\prime\prime}$-type salts according to the electron correlation.
The superconductivity of the $\beta^{\prime\prime}$-type compounds is universally expected to be produced by the charge fluctuation enhanced near the charge ordered phase.



\begin{thebibliography}{99}
\bibitem{1} C. C. Tsuei and J. R. Kirtley, Rev. Mod. Phys. {\bf72}, 969 (2000).
\bibitem{2} C. Pfleiderer, Rev. Mod. Phys. {\bf81}, 1551 (2009).
\bibitem{3} K. Kanoda, J. Phys. Soc. Jpn. {\bf75}, 051007 (2006).
\bibitem{4} H. Kontani and S. Onari, Phys. Rev. Lett. {\bf104}, 157001 (2010).
\bibitem{5} M. Tsujimoto, Y. Matsumoto, T. Tomita, A. Sakai, and S. Nakatsuji, Phys. Rev. Lett. {\bf113}, 267001 (2014).
\bibitem{6} J. Merino and R. H. McKenzie, Phys. Rev. Lett. {\bf87}, 237002 (2001).
\bibitem{7} Y. Ihara, H. Seki, and A. Kawamoto, J. Phys. Soc. Jpn. {\bf82}, 083701 (2013).
\bibitem{8} Y. Ihara, M. Jeong, H. Mayaffre, C. Berthier, M. Horvati$\rm{\acute{c}}$, H. Seki, and A. Kawamoto, Phys. Rev. B {\bf90}, 121106(R) (2014).
\bibitem{9} Y. Ihara, Y. Futami, and A. Kawamoto, J. Phys. Soc. Jpn. {\bf85}, 014601 (2016).
\bibitem{10} Y. Ihara, K. Moribe, S. Fukuoka, and A. Kawamoto, Phys. Rev. B {\bf100}, 060505(R) (2019).
\bibitem{11} W. Lubczynski, S. V. Demishev, J. Singleton, J. M. Caulfield, L. du Croo de Jongh, C. J. Kepert, S. J. Blundell, W. Hayes, M. Kurmoo, and P. Day, J. Phys.: Condens. Matter {\bf8}, 6005 (1996).
\bibitem{12} H. Mori, I. Hirabayashi, S. Tanaka, T. Mori, Y. Maruyama, and H. Inokuchi, Solid. State. Commun. {\bf80}, 411 (1991).
\bibitem{12p5} A. Girlando, M. Masino, J. A. Schlueter, N. Drichko, S. Kaiser, and M. Dressel, Phys. Rev. B {\bf89}, 174503 (2014).
\bibitem{13} A. Pustogow, Y. Saito, A. Rohwer, J. A. Schlueter, and M. Dressel, Phys. Rev. B {\bf99}, 140509(R) (2019).
\bibitem{14} A. Pustogow, K. Treptow, A. Rohwer, Y. Saito, M. Sanz Alonso, A. Lohle, J. A. Schlueter, and M. Dressel, Phys. Rev. B {\bf99}, 155144 (2019).
\bibitem{17} S. Uji, Y. Iida, S. Sugiura, T. Isono, K. Sugii, N. Kikugawa, T. Terashima, S. Yasuzuka, H. Akutsu, Y. Nakazawa, D. Graf, and P. Day, Phys. Rev. B {\bf97}, 144505 (2018).
\bibitem{18} S. Imajo, Y. Nakazawa, and K. Kindo, J. Phys. Soc. Jpn. {\bf87}, 123704 (2018).
\bibitem{16p5} T. G. Prokhorova, E. B. Yagubski, Russ. Chem. Rev. {\bf86}, 164 (2017).
\bibitem{15} A. I. Coldea, A. F. Bangura, J. Singleton, A. Ardavan, A. Akutsu-Sato, H. Akutsu, S. S. Turner, and P. Day, Phys. Rev. B {\bf69}, 085112 (2004).
\bibitem{16} A. F. Bangura, A. I. Coldea, J. Singleton, A. Ardavan, A. Akutsu-Sato, H. Akutsu, S. S. Turner, P. Day, T. Yamamoto, and K. Yakushi, Phys. Rev. B {\bf72} 014543 (2005).
\bibitem{19} H. Akutsu, A. Akutsu-Sato, S. S. Turner, D. Le Pevelen, P. Day, V. Laukhin, A.-K. Klehe, J. Singleton, D. A. Tocher, M. R. Probert, and J. A. K. Howard, J. Am. Chem. Soc. {\bf124}, 12430 (2002).
\bibitem{20} L. Martin, A. L. Morritt, J. R. Lopez, Y. Nakazawa, H. Akutsu, S. Imajo, Y. Ihara, B. Zhang, and Y. Guo, Dalton Trans. {\bf46}, 9542 (2017).
\bibitem{21} T. G. Prokhorova, L. I. Buravov, E. B. Yagubskii, L. V. Zorina, S. S. Khasanov, S. V. Simonov, R. P. Shibaeva, A. V. Korobenko and V. N. Zverev, CrystEngComm {\bf13}, 537 (2011).
\bibitem{22} E. Coronado, S. Curreli, C. Gim$\rm{\acute{e}}$nez-Saiz, and C. J. G$\rm{\acute{o}}$mez-Garc$\rm{\acute{i}}$a, Inorg. Chem. {\bf51}, 1111 (2012).
\bibitem{23} T. G. Prokhorova, L. V. Zorina, S. V. Simonov, V. N. Zverev, E. Canadell, R. P. Shibaeva and E. B. Yagubskii, CrystEngComm {\bf15}, 7048 (2013).
\bibitem{24} T. G. Prokhorova, E. B. Yagubskii, L. V. Zorina, S. V. Simonov, V. N. Zverev, R. P. Shibaeva, and L. I. Buravov, Crystals {\bf8}, 92 (2018).
\bibitem{25} T. Sasaki, N. Yoneyama, N. Kobayashi, Y. Ikemoto, and H. Kimura, Phys. Rev. Lett. {\bf92}, 227001 (2004).
\bibitem{26} T. Mori and M. Katsuhara, J. Phys. Soc. Jpn. {\bf71}, 826 (2002).
\bibitem{27} T. G. Prokhorova, L. I. Buravov, E. B. Yagubskii, L. V. Zorina, S. V. Simonov, V. N. Zverev, R. P. Shibaeva, and E. Canadell, Eur. J. Inorg. Chem. {\bf34}, 5611 (2015).
\bibitem{28} A. Audouard, V. N. Laukhin, L. Brossard, T. G. Prokhorova, E. B. Yagubskii, and E. Canadell, Phys. Rev. B {\bf69}, 144523 (2004).
\bibitem{29} S. Rashid, S. S. Turner, P. Day, J. A. K. Howard, P. Guionneau, E. J. L. McInnes, F. E. Mabbs, R. J. H. Clark, S. Firth, and T. Biggs, J. Mater. Chem. {\bf11}, 2095 (2001).
\bibitem{30} P. A. Goddard, S. W. Tozer, J. Singleton, A. Ardavan, A. Abate, and M. Kurmoo, J. Phys.: Condens. Matter {\bf14}, 7345 (2002).
\bibitem{31} H. Fukuyama and K. Yoshida, J. Phys. Soc. Jpn. {\bf46}, 102 (1979).
\bibitem{32} T. Mori, A. Kobayashi, Y. Sasaki, H. Kobayashi, G. Saito, and H. Inokuchi, Bull. Chem. Soc. Jpn. {\bf57}, 627 (1984).
\bibitem{33} D. Beckmann, S. Wanka, J. Wosnitza, J. A. Schlueter, J. M. Williams, P. G. Nixon, R. W. Winter, G. L. Gard, J. Ren, and M. -H. Whangbo, Eur. Phys. J. B {\bf1}, 295 (1998).
\bibitem{34} L. Martin, J. R. Lopez, H. Akutsu, Y. Nakazawa, and S. Imajo, Inorg. Chem. {\bf56}, 14045 (2017).
\bibitem{35} A. L. Morritt, J. R. Lopez, T. J. Blundell, E. Canadell, H. Akutsu, Y. Nakazawa, S. Imajo, and L. Martin, Inorg. Chem, {\bf58}, 10656 (2019).

\end{thebibliography}
\end{document}